\documentclass[10pt]{iopart} 

\usepackage[utf8]{inputenc}
\usepackage{iopams}
\usepackage[numbers,compress]{natbib}
\usepackage{lmodern}
\usepackage{amssymb}
\usepackage{graphicx}
\usepackage{dcolumn}                %% align table column on decimal point(s)
\usepackage{textcomp}               %% If you want to use special symbols in LaTeX, you should load 
\usepackage{subfig}

\usepackage[english]{babel}

\renewcommand{\etal}{\textit{et\ al}}
\newcommand{\text}[1]{\ensuremath{\mathrm{#1}}}     %% text style in formulas, usually in upper of lower index
\newcommand{\vect}[1]{\ensuremath{\boldsymbol{#1}}} %% vectors
\newcommand{\chem}[1]{\ensuremath{\mathrm{#1}}}     %% chemical symbols
\renewenvironment{pmatrix}%                         %% 3 by 3 (or smaller) matrix in round brackets
   {\left(\begin{array}{ccc}}%
   {\end{array}\right)}

\begin{document}

\title[Magnetic anisotropy and chirality of frustrated Cr nanostructures on Au(111)]{Magnetic anisotropy and chirality of frustrated Cr nanostructures on Au(111)}

\author{L Balogh$^1$, L Udvardi$^{1,2}$ and L Szunyogh$^{1,2}$}

\address{$^1$~Department of Theoretical Physics, Budapest University of Technology and Economics, Budafoki \'ut 8., H-1111 Budapest, Hungary}
\address{$^2$~MTA-BME Condensed Matter Research Group, Budapest University of Technology and Economics, Budafoki \'ut 8., H-1111 Budapest, Hungary}
\ead{udvardi@phy.bme.hu}

\begin{abstract}
By using a fully relativistic embedded cluster Green's function technique we investigated the magnetic anisotropy properties of four different compact Cr trimers (equilateral triangles) and Cr mono-layers deposited on Au(111) surface in both fcc and hcp stackings. 
For all trimers the magnetic ground state was found a frustrated 120\textdegree\ N\'eel configuration.  
Applying global spin rotations to the magnetic ground state, 
the predictions of an appropriate second order spin Hamiltonian 
were reproduced with high accuracy by the first principles calculations.
For the Cr trimers with adjacent Au atoms in similar geometry we obtained similar values for the in-plane
and out-of-plane anisotropy parameters, however, the Dzyaloshinskii--Moriya (DM) interactions appeared to differ remarkably. 
For two kinds of trimers we found an unconventional magnetic ground state showing 90\textdegree\ in-the-plane rotation with respect to the high symmetry directions. 
Due to higher symmetry, the in-plane anisotropy term was missing for the mono-layers and distinctly different DM interactions were  obtained for the different stackings. The chiral degeneracy of the N\'eel configurations was lifted by less then 2\;meV for the trimers, while this value raised up to about 15\;meV per 3~Cr atoms for the hcp packed mono-layer. 
\end{abstract}


%\submitto{\JPCM}

\maketitle

\section{Introduction}

Accurately describing frustrated magnetic systems both experimentally and theoretically is still a challenge.
Development in spin-polarized scanning tunnelling microscopy (SP-STM)  
made it possible to explore non-collinear magnetic structures in atomic resolution.~\cite{Bode2007-Nature,Wiesendanger2009-Review}
Frustrated non-collinear magnetic structures were reported by Gao \etal~\cite{Gao2008} for Mn islands deposited on Ag(111) surface. 
Based on topographic measurements which only collect information from the Mn layer and the topmost Ag layer, 
they concluded that fcc up and hcp down type islands and fcc stacked stripes were present.
They demonstrated that the islands exhibit the 120\textdegree\ Néel magnetic structure 
and the orientation of the Mn moments differs by 30\textdegree\ between fcc and hcp stacked islands 
most likely due to spin-orbit coupling which is different for the two stackings.
The latter phenomenon made it possible to distinguish the islands with different stackings.

\begin{figure*}[!tb]
  \subfloat[Fcc up, $\kappa^z=+1$.\label{subfig:geom-fcc-up}]{\includegraphics[angle=-90,width=0.25\textwidth]{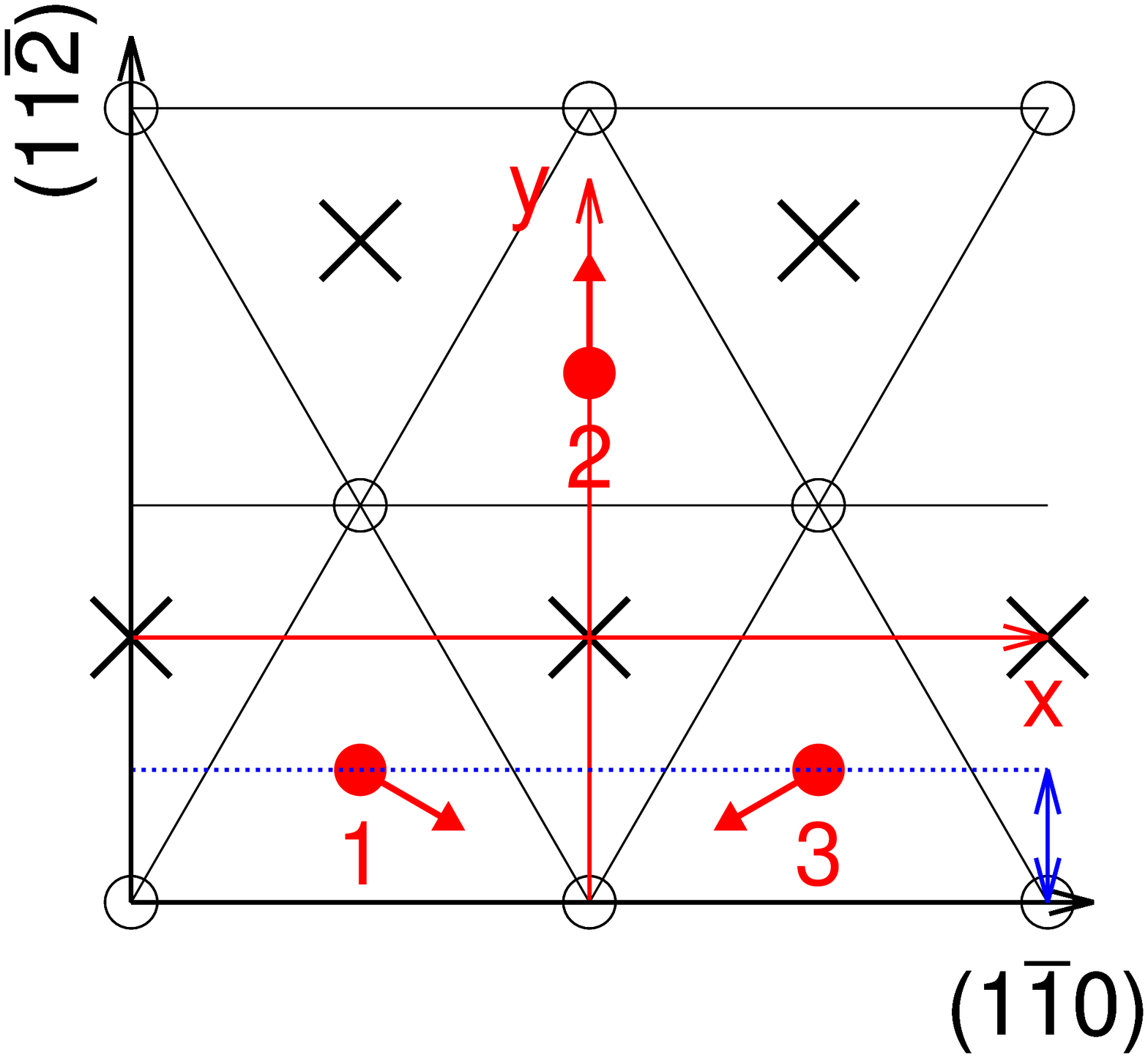}}%
  \subfloat[Fcc down, $\kappa^z=-1$.\label{subfig:geom-fcc-dn}]{\includegraphics[angle=-90,width=0.25\textwidth]{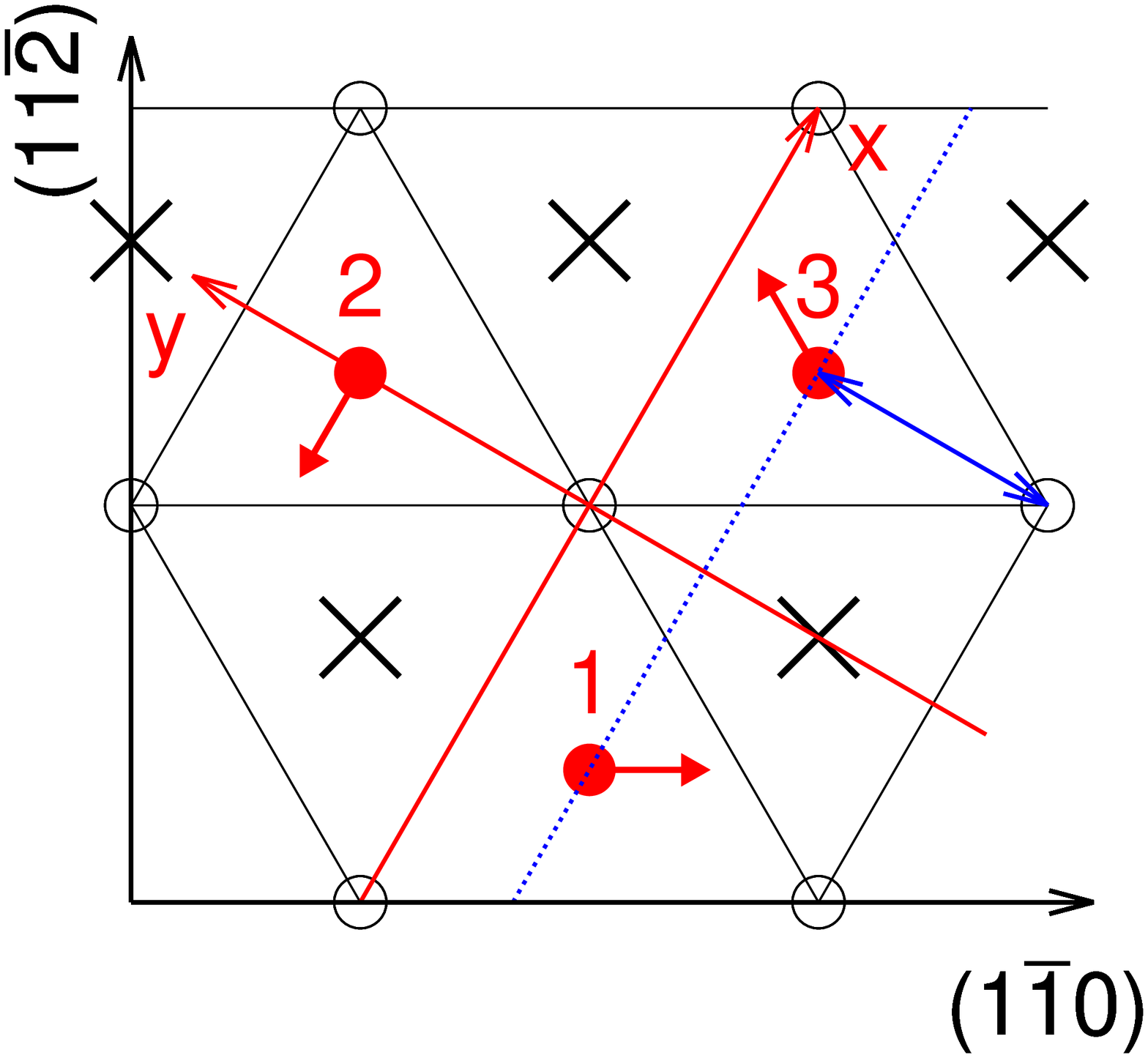}}%
  \subfloat[Hcp up, $\kappa^z=-1$.\label{subfig:geom-hcp-up}]{\includegraphics[angle=-90,width=0.25\textwidth]{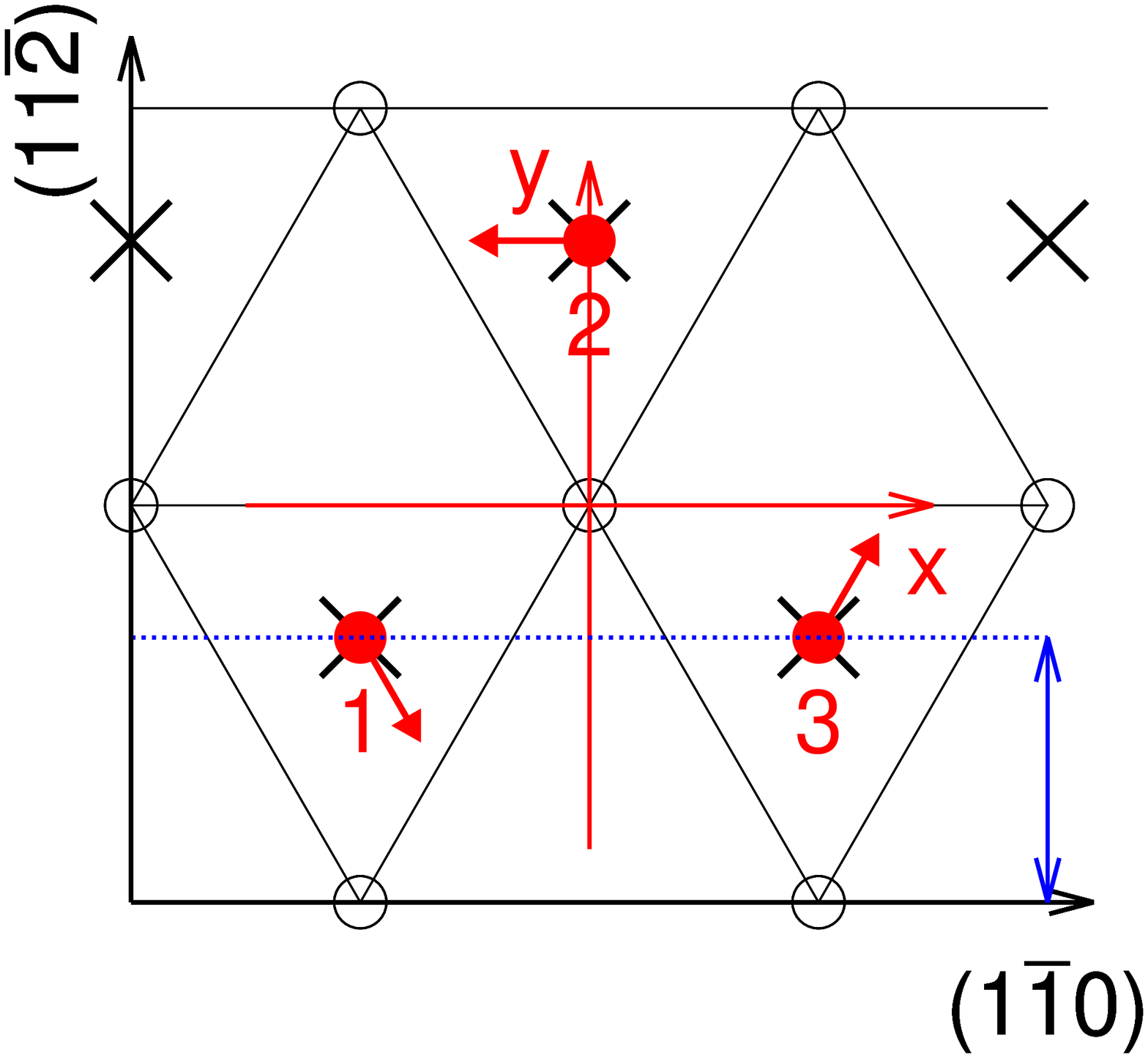}}%
  \subfloat[Hcp down, $\kappa^z=+1$.\label{subfig:geom-hcp-dn}]{\includegraphics[angle=-90,width=0.25\textwidth]{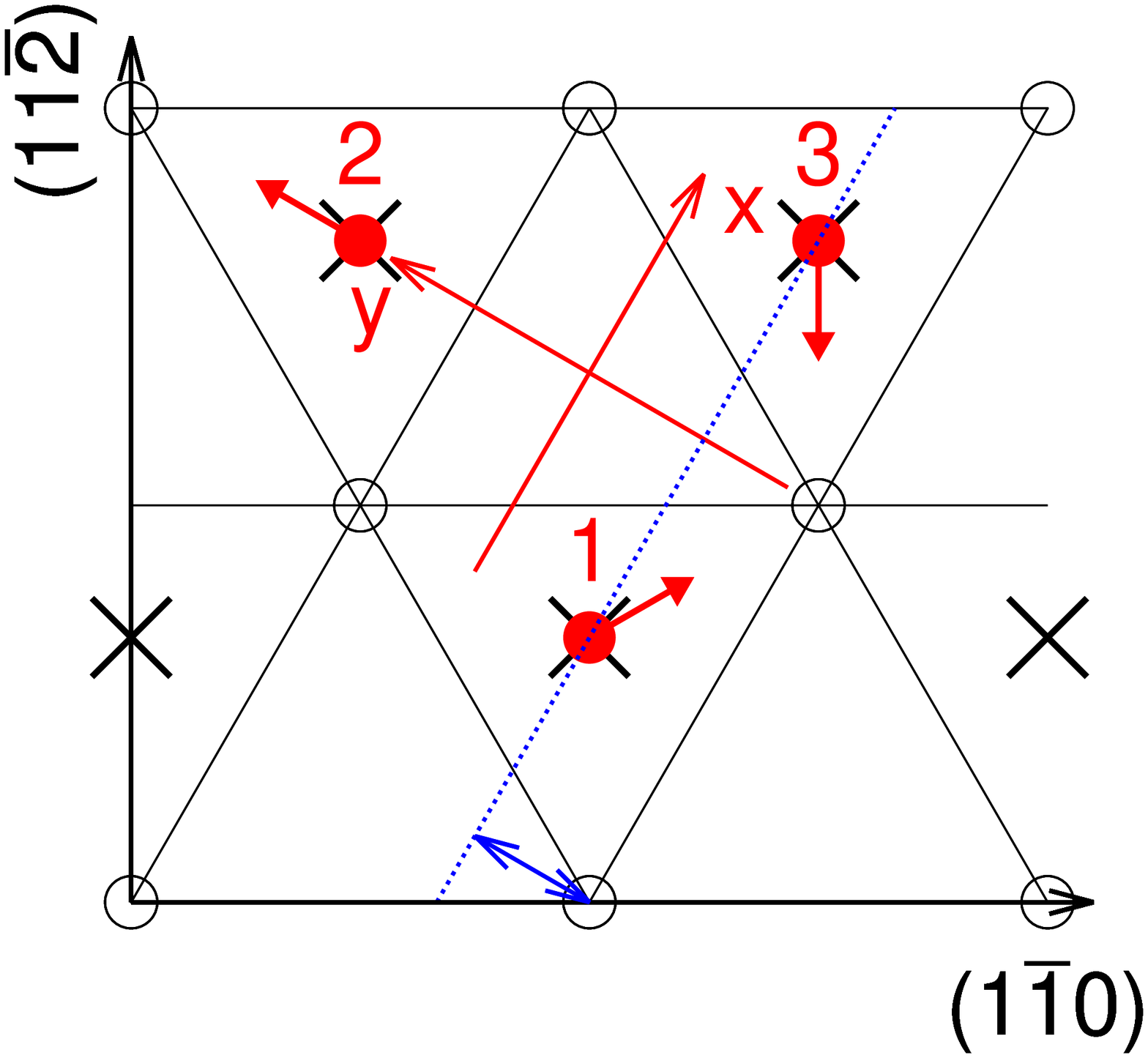}}%
  \caption{\label{fig:geometry}%
           Top view of the fcc(111) supporter and the deposited trimers.
           The atomic positions in the first (second) layer are marked with the lattice of empty circles ($\times$~signs). 
           The positions of the ad-atoms are displayed as filled red circles.
           The blue double arrows mark the two possible lateral displacements between the trimers
           and the atomic rows of the supporter measuring towards the base of the trimers. 
           For the symmetry analysis of the interaction matrices we use local coordinate systems as displayed in red colour.
           The ground state magnetic configurations are drawn by the thick red arrows. 
           The indicated chiralities in the sub-captions refer to the (\ref{eq:def:chirality})~definition.
          }
\end{figure*}

The most simple system exhibiting geometric frustration is an antiferromagnetically~(AF) coupled symmetric
trimer. An equilateral compact chromium trimer deposited on the (111) surface of gold is an
archetype of such systems. 
The first non-collinear magnetic calculations of supported metallic 3d triangular trimers were
presented fifteen years ago by Uzdin \etal.~\cite{Uzdin1999} Within the vector Anderson model they showed that a supported
equilateral Cr trimer exhibited zero net magnetic moment with the atomic moments enclosing 120\textdegree\ angles.
\emph{Ab initio} investigations of the compact Cr trimer deposited on Au(111) surface were first performed by Gotsits \etal.~\cite{Gotsits2006}
They performed spin-polarized electronic structure calculation using the projector augmented-wave
method with the spin-orbit coupling (SOC) included for optimizing the geometric and the magnetic configuration. 
Starting the optimization from fcc sites, they found that the equilateral geometry was preserved in equilibrium  
and the average of the magnetization at all the sites were laying in the plane enclosing 120\textdegree\ 
with each other.
The 120\textdegree\ Néel state was also found to be the lowest energy magnetic configuration of
the Cr trimer supported on a Au(111) surface by Bergman \etal~\cite{Bergman2007-CondMat} using an extension of the real space linear
muffin tin orbital method (RS-LMTO) within the atomic sphere approximation (ASA) with relativistic effects included within the scalar relativistic approximation.

Using fully relativistic constrained self-consistent multiple scattering Green's function electronic structure
calculations~\cite{Stocks1999} and adiabatic spin dynamics~\cite{Antropov1995} to search for the ground
state, Stocks \etal~\cite{Stocks2007} also concluded that the ground state of the compact Cr trimer on Au(111) is the
120\textdegree\ Néel state. The energy of the two magnetic configurations with opposite
chirality turned out to be different as it has also been confirmed by Antal \etal.~\cite{Antal2008} 
They pointed out that the  Dzyaloshinskii--Moriya (DM) interaction is responsible for the lifting
of the degeneracy of the states with opposite chirality.

The 120\textdegree\ Néel state of Cr mono-layer on Pd(111) substrate was observed by Wa\'{s}niowska \etal~\cite{Wasniowska2010} by SP-STM measurements.
From first-principles calculations on flat spin spirals, they predicted the ground state to be the
120\textdegree\ Néel state which is in agreement with the measurements. 
They found the Cr ML to be energetically favourable in fcc than in hcp stacking by about $162\;\mathrm{meV}/\text{Cr~atom}$.
Palot\'as \etal~\cite{Palotas2011} demonstrated by first principles
calculations that the two possible N\'eel states with opposite chiralities
of a Cr mono-layer on Ag(111) are energetically in-equivalent.
They also showed that the magnetic contrast of the simulated SP-STM
image was sensitive to the electronic structure of the tip and
to the bias voltage.

The aim of the present work is to systematically investigate the magnetic anisotropy of the four Cr trimers and to compare them to those of Cr mono-layers deposited on Au(111). 
After presenting the geometrical structure of the clusters and mono-layers we briefly describe the applied methods used to determine the electronic and magnetic structure of the systems. 
By exploiting the point group symmetry of the systems, 
a suitable classical Heisenberg model is constructed which is then applied to 
give analytic forms of the rotational energies. 
These forms are used to analyze and discuss the results of the ab-initio calculations 
for the four different trimers and for the mono-layers with fcc and hcp stacking. 
The chirality of the trimers and the mono-layers is discussed and the values of the accessible model parameters are given.

\section{Details of the calculations}

The high symmetry adsorption sites of a fcc(111) surface are the fcc hollow, the hcp hollow, the bridge and 
the on-top sites.
The on-top positions of the Cr on Au(111) is energetically unfavourable as it was pointed out by
Gotsits \etal~\cite{Gotsits2006}.
For trimers occupying the bridge positions the bond length would be close to the half of the lattice
constant of the underlying supporter which is too short compared to the relaxed bond length predicted by
ab-initio calculations~\cite{Gotsits2006}. Considering these facts, in the present paper we investigate 
fcc and hcp stacked compact trimers located only at the hollow sites of the substrate.

Equilateral trimers can be deposited in the hollow positions of the fcc(111) surface in four
different configurations as it is shown in figure~\ref{fig:geometry}.
By labelling the in-equivalent layers of the fcc lattice along the [111] direction with capital letters, the order of the 
fcc stacking is \mbox{ABCABC\textbf{A}}, while the order for the hcp stacking is \mbox{ABCABC\textbf{B}},
where the last (boldface) symbol corresponds to the deposited trimers or mono-layers.  
Fcc and hcp stacked trimers both can be either \emph{up} or \emph{down} triangles, see figure~\ref{fig:geometry}, 
which can be distinguished by the lateral displacement between the supporter atomic rows and the base of the triangle as it is indicated by the blue double arrows in figure~\ref{fig:geometry}.
Note, that one cannot distinguish between an fcc up and an hcp down island (fcc down and hcp up) 
if one only sees the island itself and the topmost supporter layer which is the situation in 
a constant current non-magnetic STM experiment.

Previous studies on Cr clusters forming equilateral triangle considered the fcc up trimer~\cite{Gotsits2006,Bergman2007-CondMat,Stocks2007,Antal2008} 
and, according to our knowledge, magnetic properties of the other compact Cr trimers have not been investigated yet.
The four clusters are, however, different: the Cr atoms in an fcc up or an hcp down
cluster surround an interstice in the first supporter layer (\emph{breezy} triangle) 
while in an fcc down or an hcp up cluster they surround a substrate atom (\emph{crammed} triangle).
The trimers with fcc and hcp stacking together with the substrate exhibit a $C_{3v}$ point group symmetry
where the $C_3$ axis intersects the centre of the triangle normal to the substrate
and the reflection planes contain the $C_3$ axis and one of the cluster atom.
For the mono-layers with fcc and hcp stacking the $C_{3v}$ point group symmetry also holds 
with $C_3$ axes intersecting the centre of an elementary triangle or a Cr atom. 
Note that the mono-layers contain alternating up and down triangles.
The two types of triangles are in-equivalent, i.e., one type of
triangle can not be transformed into the other type by any of the symmetry operations of the system.

The electronic structure of the Cr mono-layers and the Cr trimers on top of
Au(111) surface was calculated in terms of the fully relativistic screened KKR 
method~\cite{Szunyogh1994-PRB,Szunyogh1994-JPCM} within the local spin-density approximation 
(LSDA),~\cite{Ceperley--Alder1980} of the density functional theory as parametrized by Perdew and 
Zunger~\cite{Perdew--Zunger1981}. In particular, for the clusters  we applied the embedded 
cluster Green's function technique~\cite{Lazarovits2002} based on the KKR method.
The effective potentials and fields were treated within the atomic sphere approximation (ASA),
and a cutoff of $\ell_\text{max}=2$ for the angular momentum expansion was used.
For the case study presented in this work we neglected the relaxation of the layer-layer distances and the 2D lattice constant of gold 
($a_\text{2D}=2.874$\;\AA) was applied in all calculations.
Correspondingly, in case of the trimers the three Cr atoms occupied hollow positions above the topmost Au layer. In order to let the electron density relax around the cluster
the first neighbour shell was also included in the embedded cluster.
The magnetic ground state configurations for the trimers were determined self-consistently by using the procedure described in reference~\citenum{Balogh2012}.
The obtained directions of the magnetic moments at the Cr atoms are depicted in figure~\ref{fig:geometry}.

\begin{figure}[!tb]
 \begin{center}\includegraphics[angle=-90,width=\columnwidth]{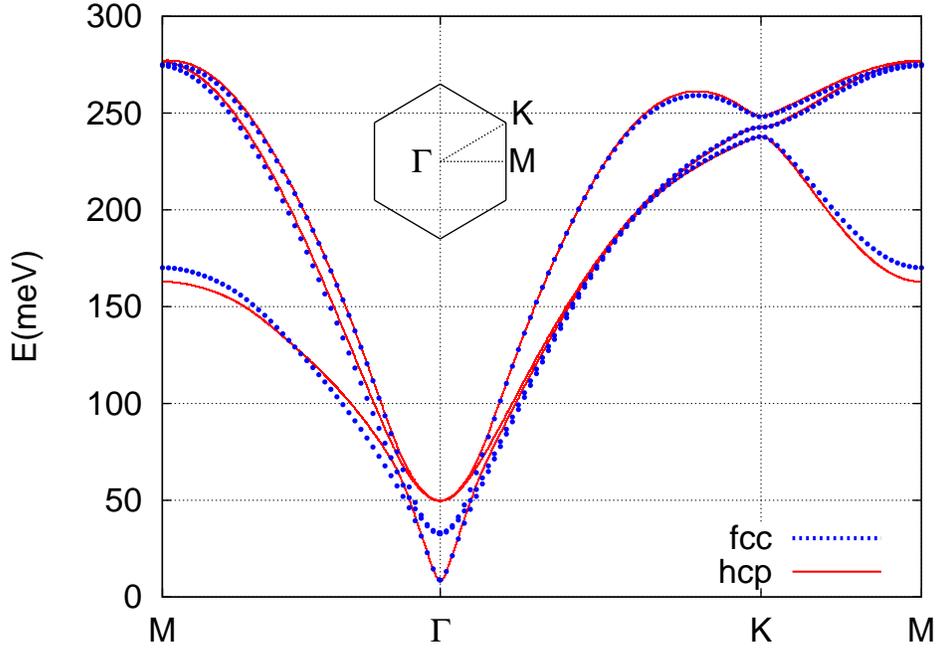}\end{center}
 \caption{Spin-excitation spectra with a reference of a 120\textdegree\ N\'eel structure of a Cr mono-layer on Au(111) with hcp and fcc stacking.
          The inset displays the special points of the hexagonal surface Brillouin zone.
          \label{fig:magnon}}
\end{figure}

The direction of the exchange field on the three Cr sub-lattices in the mono-layers were fixed according to figure~\ref{fig:geometry}(a) in the case of fcc stacking,
while according to figure~\ref{fig:geometry}(d) in the case of hcp stacking. 
For very low coverage of Cr on Au(111) surface Boeglin \etal\ reported a spin magnetic moment 
of $4.5\pm0.4\;\mu_\text{B}/\text{atom}$ for single ad-atoms while the average spin magnetic moment rapidly 
vanished with increasing island size indicating antiferromagnetic arrangement, however, 
the details of the AF order has not been explored.~\cite{Boeglin2005}
The most probable configuration is a 
120\textdegree\ N\'eel structure, however, spin structures like row-wise or double-row-wise antiferromagnetic 
alignments~\cite{Kruger2000}, or even a three-dimensional spin structure~\cite{Kurz2001} have been predicted for 2D triangular lattice.
Since our interest is focused on the magnetic anisotropy properties of a 
120\textdegree\ N\'eel structure we haven't attempted to explore the energetics of the different magnetic 
configurations. In order to confirm that the 120\textdegree\ N\'eel structure represents a local minimum of
the energy, we investigated the spin-excitation spectrum of the mono-layer with both stackings. 
The details of the calculation is given in the Appendix.
The spectrum along the special directions of the first Brillouin zone is depicted in figure~\ref{fig:magnon}. 
From our calculations we obtained no imaginary frequencies indicating that the reference magnetic configuration, i.e., the
120\textdegree\ N\'eel structure is not a meta-stable state.

\section{Spin model}

The magnetic properties of thin films and clusters of transition metals can often be successfully described by a classical Heisenberg model
\begin{equation}
 \label{eq:2nd-order-Heis}
 \mathcal{H} = \sum_{\left< i,j \right>}{\vect{\sigma}_i^\mathrm{T} \mathbf{J}_{ij} \vect{\sigma}_j} + 
               \sum_{i}{\vect{\sigma}_i^\mathrm{T} \mathbf{K}_{i} \vect{\sigma}_i}
\end{equation}
where the first summation runs over the interacting pairs of spins, $\vect{\sigma}_i$ is a unit vector parallel to the magnetization at site $i$, $\mathbf{J}_{ij}$ are generalized exchange
interaction matrices and $\mathbf{K}_i$ represents the second-order on-site anisotropy matrix.
In case of mono-layers with N\'eel spin-structure 
$\vect{\sigma}_i$ are sorted out according to the three sub-lattices.
Without loss of generality, the $\mathbf{K}_i$ matrices are chosen to be symmetric and
traceless. The $\mathbf{J}_{ij}$ exchange matrices can be decomposed into an isotropic part of $J_{ij}=\frac{1}{3}\mathrm{Tr}\,\mathbf{J}_{ij}$, a
traceless symmetric anisotropic part defined as $\mathbf{J}_{ij}^\text{S}=\frac{1}{2}\left(\mathbf{J}_{ij}+\mathbf{J}_{ij}^\mathrm{T}\right)-J_{ij}\mathbf{I}$ 
and an antisymmetric part given by $\mathbf{J}_{ij}^\text{A}=\frac{1}{2}\left(\mathbf{J}_{ij}-\mathbf{J}_{ij}^\mathrm{T}\right)$.
The latter term is usually formulated with the Dzyaloshinskii--Moriya vector, $\vect{D}_{ij}$, as 
$\vect{\sigma}_{i}^\mathrm{T}\mathbf{J}_{ij}^\text{A}\vect{\sigma}_j=\vect{D}_{ij}\left(\vect{\sigma}_i\times\vect{\sigma}_j\right)$.

We start our study of the magnetic anisotropy by symmetry considerations regarded the 
spin-Hamiltonian~(\ref{eq:2nd-order-Heis}) for the trimers.
Since the exchange tensors, $\mathbf{J}_{13}$, $\mathbf{J}_{32}$ and $\mathbf{J}_{21}$, as well as
the on-site anisotropy matrices, $\mathbf{K}_{1}$, $\mathbf{K}_{2}$ and $\mathbf{K}_{3}$, are related 
in terms of appropriate similarity transformations it is sufficient to explore the structure of one of them.
Considering the trimers in figure~\ref{fig:geometry} with respect to the local coordinate systems,
$\mathbf{J}_{13}$ and $\mathbf{J}_{31}$ are obviously related to each other as
\begin{equation}
 \mathbf{J}_{31}=\left(R^{yz}\right)^\mathrm{T}{\mathbf{J}_{13}}R^{yz} 
 =\left(R^{yz}\right)^\mathrm{T}{\mathbf{J}_{31}}^\mathrm{T}R^{yz}\,,
\end{equation}
where $R^{yz}$ denotes the reflection to the $yz$ plane.
The anisotropy matrix $\mathbf{K}_2$ must be invariant under the same operation. 
The matrices satisfying the above requirements are parametrized as:
\begin{equation}
  \label{eq:J31}
  \mathbf{J}_{31} = \begin{pmatrix}
                      J {-} \frac{1}{2}S^{zz} {-} S^{\varphi\varphi} &              D^{z}                             &   -D^{y}     \\
                                -D^{z}                               & J {-} \frac{1}{2}S^{zz} {+} S^{\varphi\varphi} &    S^{yz}    \\
                                 D^{y}                               &              S^{yz}                            &  J {+} S^{zz}
                    \end{pmatrix}
\end{equation}
and%
\begin{equation}
  \label{eq:K2}
   \mathbf{K}_{2}  = \begin{pmatrix}
                       -\frac{1}{2}K^{zz} {-} K^{\varphi\varphi} &  0                                        & 0      \\
                       0                                         & -\frac{1}{2}K^{zz} {+} K^{\varphi\varphi} & K^{yz} \\
                       0                                         &  K^{yz}                                   & K^{zz}
                    \end{pmatrix}\,.
\end{equation}

For mono-layers, the coupling between the sub-lattices $a$ and $b$, $\mathbf{J}_{ab}$, is defined as the sum
of the corresponding exchange tensors: 
$\mathbf{J}_{ab} = \sum_{j\in b} \mathbf{J}_{ij}$ for any $i\in a$. If all the sites are located in the
same sub-lattice, $\mathbf{J}_{aa}$ will contribute to the $\mathbf{K}_a$ anisotropy matrix, which will be the same for all sub-lattices.
Due to the $C_{3v}$ symmetry of the mono-layer, it turns out that the terms $K^{\varphi\varphi}$ and $ K^{yz}$ 
in equation~(\ref{eq:K2}) disappear.

Following the method proposed by Szu\-nyogh \etal~\cite{Szunyogh2009} to analyze the magnetic anisotropy
of \chem{IrMn} and \chem{IrMn_3}
the energies of the systems are calculated during the simultaneous rotation of the antiferromagnetic
configuration around the three fold axis and around an axis parallel to the magnetization on one of
the sites or sub-lattices. In the following this energy will be referred to as \emph{rotational energy}. Similar
procedure has been applied to study the magnetic anisotropy for \chem{IrMn_3/Co} interface~\cite{Szunyogh2011} and for a  cobalt
nanocontact~\cite{Balogh2012}.

The two type of 120\textdegree\ N\'eel structures can be distinguished by investigating the chirality of 
the configurations. According to Antal \etal~\cite{Antal2008} the chirality vector for a trimer is defined as
\begin{equation}
  \label{eq:def:chirality}
  \vect{\kappa} = \frac{2}{3\sqrt{3}} 
  \left( \vect{\sigma}_1 \times \vect{\sigma}_2 + \vect{\sigma}_2 \times \vect{\sigma}_3 
  + \vect{\sigma}_3 \times \vect{\sigma}_1 \right).
\end{equation}
For in-plane spin-configurations, the $\vect{\kappa}$ vector is parallel to the $z$ axis and 
its $z$ component, $\kappa^z$, being either $+1$ or $-1$ characterizes the spin-configurations.
The chiralities of the obtained spin-configurations for the trimers are also given in figure~\ref{fig:geometry}.

In the case of an in-plane Néel spin structure of a mono-layer, the chirality for the up and down triangles alternates between the values of $+1$ or $-1$.
In this case, we associate the \emph{chirality of the mono-layer} with the chirality of the \emph{up} triangles.
The energy of the Néel structures with opposite chiralities may be different due to the 
non-vanishing $z$~component of the DM vector.
While rotation around the three fold axis does not alter the chirality of an in plane N\'eel configuration,
a 180\textdegree\ rotation around an axis parallel to the magnetization on one of the sites will reverse its value.

The rotational energies were calculated in the spirit of the magnetic force
theorem~\cite{Jansen1999}. The effective potentials and exchange fields determined 
in ground state configurations were kept fixed and the change in energy of the system with respect to the
rotational angle is approximated by the change in band energy:
\begin{equation} \label{eq:Eb}
  E_\text{b}\big( \{\vect\sigma_i\} \big) 
      = \!\!\int\limits_{-\infty}^{\varepsilon_\text{F}}\!\! 
        \left( \varepsilon-\varepsilon_\text{F} \right) n\big( \varepsilon , \{\vect\sigma_i\} \big)\,\mathrm{d}\varepsilon
      = -\!\!\int\limits_{-\infty}^{\varepsilon_\text{F}}\!\! N\big( \varepsilon , \{\vect\sigma_i\} \big) \,\mathrm{d}\varepsilon 
\end{equation}
where $\varepsilon_\text{F}$ is the Fermi energy, $n\big(\varepsilon,\{\vect\sigma_i\}\big)$ 
and $N\big(\varepsilon,\{\vect\sigma_i\}\big)$ stand for the density of states (DOS) and for the integrated
DOS, respectively, and $\{\vect\sigma_i\}$ indicates the dependence of these quantities 
on the spin configuration of the system.
For the trimers, the band energy was calculated by using the Lloyd's formula~\cite{Lloyd1967}:
\begin{equation} \label{eq:Eb-Lloyd}
 E_\text{b}\big( \{\vect\sigma_i\} \big)  = -\!\!\int\limits_{-\infty}^{\varepsilon_\text{F}}\!\!
  \ln \det \left[\boldsymbol{1} + \boldsymbol{\tau}_h\left (\mathbf{t}_\mathcal{C}\big( \{\vect\sigma_i\} \big) ^{-1} - \mathbf{t}_h^{-1} \right ) \right]  \mathrm{d}\varepsilon \, ,
\end{equation}
where $\mathbf{t}_h$ and $\boldsymbol{\tau}_h$
denote the single-site scattering matrix and the scattering path operator (SPO) for the
host confined to the sites in the cluster, $\mathcal{C}$, respectively, while
$\mathbf{t}_{\mathcal{C}}$ denotes the single-site scattering matrices of
the embedded atoms~\cite{Lazarovits2002} and we omitted a constant shift of the energy  
not affected by the magnetic configuration. 
Note that using formula~(\ref{eq:Eb-Lloyd}), the change in the band energy due to the change 
of the magnetic configuration in cluster $\mathcal{C}$ is accounted for the whole system, 
while the direct integration of the local DOS in equation~(\ref{eq:Eb}) is always restricted 
to a given environment of $\mathcal{C}$ only.
To calculate the SPO of the layered host systems, the energy integrations were performed by sampling 16~points on a semicircular path in the upper complex semi-plane and
3300~$k$-points were used in the irreducible wedge of the surface Brillouin zone (SBZ).
We have checked the accuracy of the SBZ-integrals by performing the same calculations using 1900~$k$-points in the irreducible wedge of the SBZ
and a deviation of up to 4\;\% was found in the resulting model parameters. 
In the case of the mono-layers we used a fine adaptive $k$-set ranging from 
7320~$k$-points in the SBZ at the Fermi energy to 840~$k$-points at the bottom of the band
and we note that the calculations were carried out with three atoms per unit cell, 
i.e., the magnetic unit cell of the Néel structure.

Using the parametrization given by equations~(\ref{eq:J31}) and~(\ref{eq:K2}) simple expressions can be derived 
for the rotational energies based on the Heisenberg model.
For the rotations around the three fold axis the energy has the form of
\begin{eqnarray}
  \label{eq:energy-rotz+}
  E^+_z(\varphi) &&= E_{0z},\\
  \label{eq:energy-rotz-}
  E^-_z(\varphi) &&= E_{0z} - 3\sqrt{3}D^z + 
    3 \big( S^{\varphi\varphi} + K^{\varphi\varphi} \big) \cos(2\varphi),
\end{eqnarray}
where $E_{0z}$ is an energy independent on the angle of rotation and the $\pm$ superscripts indicate
whether a configuration with positive or negative chirality is rotated rigidly around the axis.
Similarly, when the configuration is rotated around the axis parallel to the magnetization
at the 2\textsuperscript{nd} Cr atom or at the 2\textsuperscript{nd} sub-lattice the energy can be given as:
\begin{eqnarray}
  \label{eq:energy-roty}
  \fl E^\pm_y(\vartheta) = E_{0y} 
       \pm \left[ \frac{3\sqrt{3}}{2}D^z - \frac{3}{2} \big( S^{\varphi\varphi} + K^{\varphi\varphi} \big) \right] \cos(\vartheta) \nonumber\\
         + \left[ \frac{3}{8} \big( S^{\varphi\varphi} + K^{\varphi\varphi} \big) + \frac{9}{16} \big( S^{zz} -2K^{zz} \big) \right] \cos(2\vartheta).
\end{eqnarray}
The $\pm$~superscript indicates here the chirality at $\vartheta=0$.
We note again that the rotation around the $y$ axis reverses the chirality but the rotation around 
the $z$~axis does not alter the chirality.

By comparing these rotational energy functions to the results of the first principles calculations the values of the coefficients of the trigonometric functions can be extracted. 
From equation~(\ref{eq:energy-roty}) the energy difference between the positive and the negative chirality configuration can be read off:
$\Delta{}E=E^{+}_{y}(0)-E^{-}_{y}(0)=3\sqrt{3}D^z-3(S^{\varphi\varphi}+K^{\varphi\varphi})$.

Since the relative angle between the spins does not change during the global spin rotations,
the contribution of the isotropic exchange cancel out and the $J$ parameter of the model is not accessible through the $E^\pm_{z}(\varphi)$ and $E^\pm_{y}(\vartheta)$ functions. 
Regarding the focus of the recent work, the factual value of the isotropic exchange is, therefore, irrelevant. 
Furthermore, we note that the rotational invariant fourth order terms introduced by Antal \etal~\cite{Antal2008} neither contribute to the rotational energy.

\begin{figure}[!tb]
  \begin{center}\includegraphics[angle=-90,width=\columnwidth]{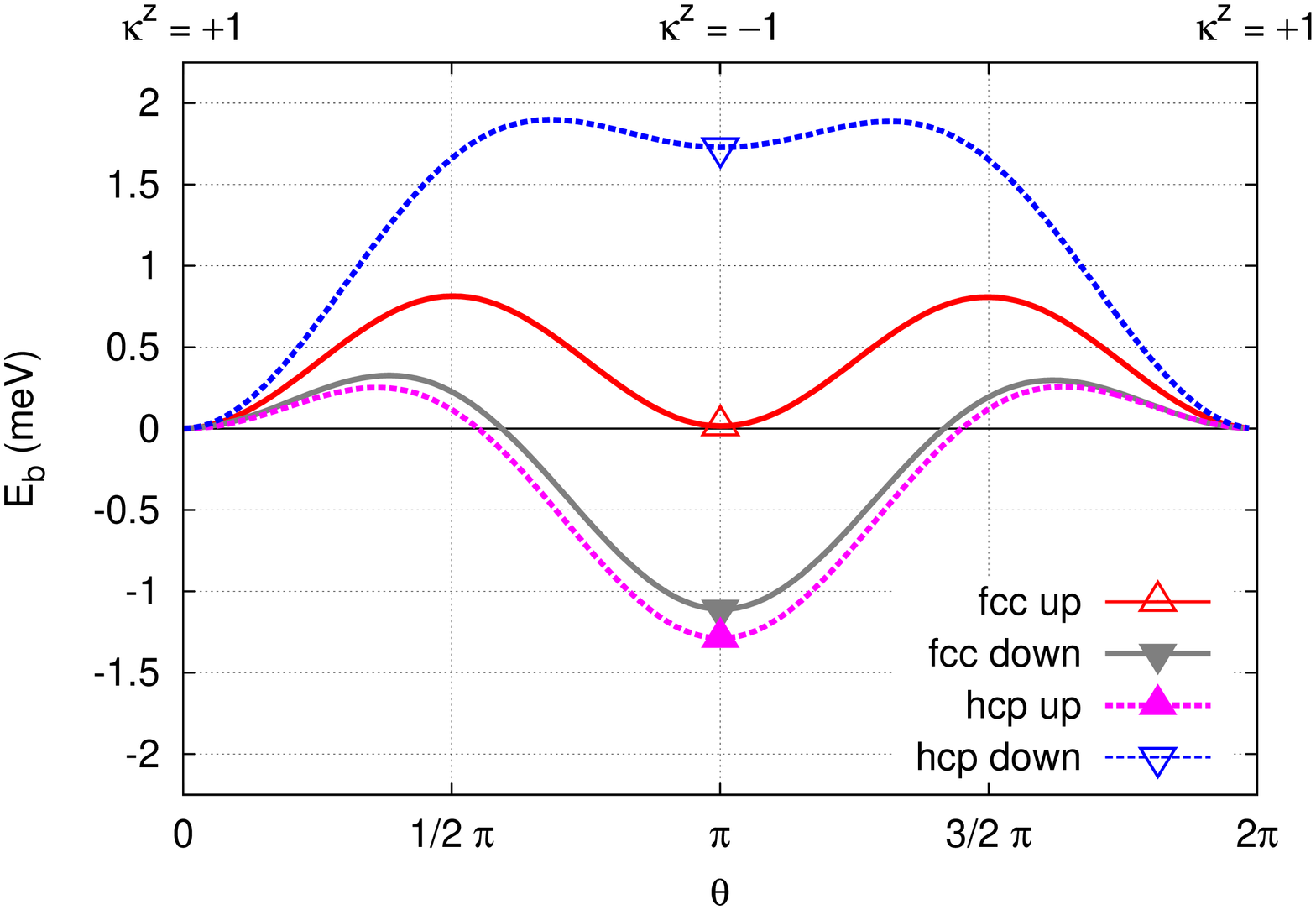}
                \includegraphics[angle=-90,width=\columnwidth]{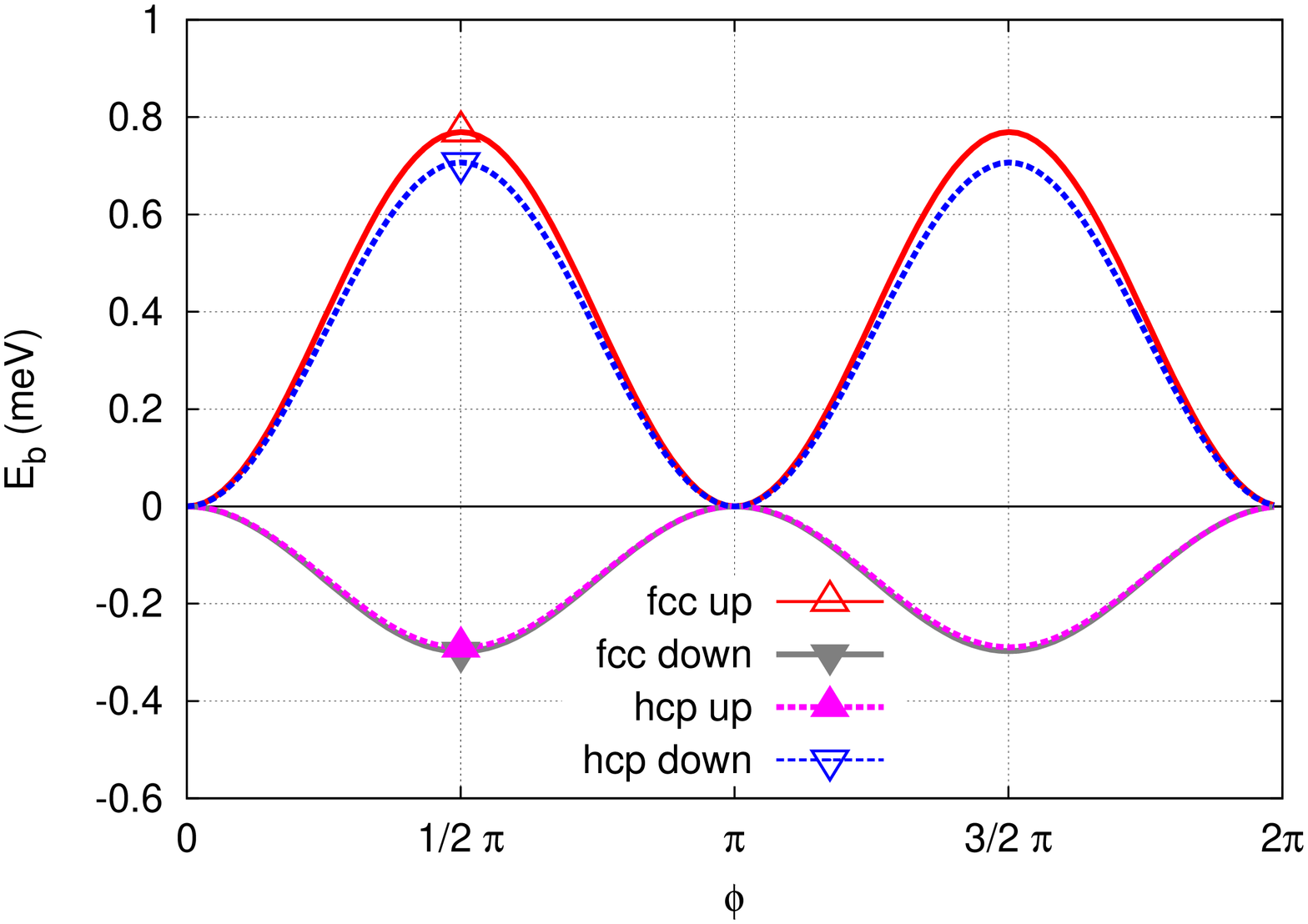}\end{center}
  \caption[]{Rotational energy of the trimers,
             $E^{+}_{y}(\vartheta)$ (upper graph) and  $E^{-}_{z}(\varphi)$ (lower graph). 
             The symbols refer to the trimer geometries shown in figure~\ref{fig:geometry}.
             The points forming the lines were calculated with a resolution of~3\textdegree.
             The $\kappa^z$ component of the chirality vector is indicated above the upper graph 
             and we note that a global rotation about the $z$ axis (lower graph) does not alter the chirality.
             The energy curves are shifted to zero at the initial configuration, $\vartheta=0$ or $\varphi=0$.
             \label{fig:Erotyrotz}
             }
\end{figure}

\section{Trimers}

In the case of the trimers the spin moments of the Cr atoms scattered between $4.18\;\mu_\text{B}$ and $4.22\;\mu_\text{B}$ 
while the orbital moments between $0.021\;\mu_\text{B}$ and $0.038\;\mu_\text{B}$ 
depending on the geometry and the magnetic structure. 
These values are in good agreement with the results of previous studies:
$3.15\;\mu_\text{B}/\text{atom}$ if geometrical relaxation is included~\cite{Gotsits2006}; 
$4.25\;\mu_\text{B}/\text{atom}$~\cite{Bergman2007-CondMat} and 
$4.4\;\mu_\text{B}/\text{atom}$~\cite{Antal2008} if it is neglected. 
For the orbital moment also small values have been reported: 
${\leq}0.036\;\mu_\text{B}/\text{atom}$ with geometrical relaxation included~\cite{Gotsits2006} and 
${\approx}0.03\;\mu_\text{B}/\text{atom}$~\cite{Antal2008} without relaxation.

The rotational energies are calculated with a resolution of 3\textdegree\ for all four trimers and the results are shown in figure~\ref{fig:Erotyrotz}.
The parameters in equations~(\ref{eq:energy-rotz-}) and~(\ref{eq:energy-roty}) are obtained as the Fourier components of the rotational energies and listed in table~\ref{tab:parameters}.
Using these parameters the functions given by equations~(\ref{eq:energy-rotz-}) and~(\ref{eq:energy-roty})
fit with a high accuracy to the results provided by the ab-initio calculations.

\begin{table}[!tb]
  \begin{indented}
  \item[]
  \caption[]{Fitted spin-model parameters entering equations~(\ref{eq:energy-rotz-}) and~(\ref{eq:energy-roty})
  together with the energy difference between the two chiral states, $\Delta E=E^+_y(0)-E^-_y(0)$, for the four different trimers and the two mono-layers~(ML).
             The values are given in units of meV.
             \label{tab:parameters}
             }
  \begin{tabular}{lD{.}{.}{3}D{.}{.}{3}D{.}{.}{3}D{.}{.}{3}}
  \br
  Trimer/ML & \multicolumn{1}{c}{$S^{\varphi\varphi}+K^{\varphi\varphi}$} 
                                    & \multicolumn{1}{c}{$D^z$} & \multicolumn{1}{c}{$S^{zz}-2K^{zz}$} & \multicolumn{1}{c}{$\Delta E$} \\
  \mr
  fcc up    &                -0.128 &                    -0.077 &                               -0.629 &                                                   -0.018 \\
  fcc down  &                 0.050 &                     0.242 &                               -0.714 &                                                   +1.110 \\
  hcp up    &                 0.048 &                     0.276 &                               -0.711 &                                                   +1.288 \\
  hcp down  &                -0.118 &                    -0.401 &                               -0.625 &                                                   -1.730 \\
  \mr
  fcc ML    & {}<1.\;\mu\mathrm{eV} &                    -1.086 &                               -0.469 &                                                   -5.643 \\
  hcp ML    & {}<1.\;\mu\mathrm{eV} &                     2.972 &                               -0.482 &                                                  +15.444 \\
  \br
  \end{tabular}
  \end{indented}
\end{table}

From figure~\ref{fig:Erotyrotz} it can be inferred that the energy minimum corresponds to $\kappa^z=+1$
for the fcc up and hcp down trimers, while to $\kappa^z=-1$ for the fcc down and hcp up trimers, see also figure~\ref{fig:geometry}.
Note, however, that in case of the fcc up trimer the energy difference between the two chiral states is found to be $-18\;\mu$eV
which is near the computational accuracy of our method.

In the case of the $\kappa^z=-1$, the relative orientation of the magnetization vector and
the easy direction set by the on-site anisotropy term ($\mathbf{K}_i$) is the same for
the three Cr atoms and this situation is preserved during the global in-plane rotation,
therefore, the anisotropy energies of the single atoms are simply summed up. 
The same argument holds for the two-site anisotropies ($\mathbf{J}^{\text{S}}_{ij}$), 
thus, as indicated by equation~(\ref{eq:energy-rotz-}), we expect a
$\cos(2\varphi)$ angular dependence for the in-plane rotational energy. 
This is clearly confirmed by the first principles calculations, see the lower graph of figure~\ref{fig:Erotyrotz}.

For the in-plane rotational energy of the $\kappa^z=+1$ trimers we expect the anisotropy terms to cancel since 
the second order in-plane anisotropy energies are sampled at angles $\varphi_1=\varphi$, $\varphi_2=120^\circ+\varphi$, and $\varphi_3=240^\circ+\varphi$, for which
$\sum_{i=1}^{3}\cos^2\varphi_i=\frac{3}{2}$, i.e., independent of the angle of rotation.   
The magnitude of the rotational energies of the positive chirality trimers was indeed found below 7\;$\mu$eV,
indicating a very small deviation between the spin model~(\ref{eq:2nd-order-Heis}) and the \emph{ab initio} calculation.
Similarly, Szu\-nyogh \etal~\cite{Szunyogh2009} found a $\cos(2\varphi)$ angular dependence of the rotational energy of \chem{IrMn_3} with an amplitude of $10.42\;\mathrm{meV}$ in the so-called 
$T1$ state with negative chirality, while for the states with positive chirality the rotational energy had no angular dependence up to an absolute error of $2\;\mu\mathrm{eV}$.

It can be read off from table~\ref{tab:parameters} that the trimers with similar environment, i.e., the breezy and the crammed triangles, exhibit similar parameter values. 
This is, in particular, valid for the out-of-plane and in-plane anisotropy parameters, $S^{zz}-2K^{zz}$ and $S^{\varphi\varphi}+K^{\varphi\varphi}$, respectively. 
The $z$ component of the DM vector turned out to be similar for the fcc down and the hcp up (crammed) trimers,
but $D^z$ for the fcc up and the hcp down (breezy) trimers are rather different.
We notice that for an fcc up trimer Antal \etal~\cite{Antal2008} reported a value of $D^z=0.97\;\mathrm{meV}$ which is a remarkable difference compared to our present
value of $D^z=-0.077\;\mathrm{meV}$.
There are, however, distinct differences between the two calculations. On the one hand,
here we included one shell of environment around the atoms forming the trimer, whereas in reference~\citenum{Antal2008} only the Cr atoms were taken into
account in the self-consistent calculations. On the the other hand, we calculated the rotational energies in terms of the Lloyd's formula, equation~(\ref{eq:Eb-Lloyd}), while Antal \etal~\cite{Antal2008}
used equation~(\ref{eq:Eb}) to evaluate the band-energy.

In case of the fcc up trimer, we repeated the magnetic force theorem calculation 
of the rotational energies by using the self-consistent effective potentials and fields 
from the negative chirality configuration and found that even in this case the positive chirality state was lower in energy.
Remarkably, however, Stocks \etal~\cite{Stocks2007} found considerable difference for $\Delta{}E=E^{+}_{y}(0)-E^{-}_{y}(0)$ if they used the (negative chirality) Néel state ($\Delta{}E=+7\;\mathrm{meV}$) 
or the out-of-plane ferromagnetic state ($\Delta{}E=-4\;\mathrm{meV}$) for the self-consistent reference potential and field calculation.

Regarding the in-plane anisotropy (see the lower graph of figure~\ref{fig:Erotyrotz}),
for the fcc up and hcp down trimers we found a value of $S^{\varphi\varphi}+K^{\varphi\varphi}$ which is about 50\;\% larger in magnitude as compared
to the fcc up trimer calculations of Stocks \etal~\cite{Stocks2007}.
The reason for this difference is the same as mentioned above in context to $D^z$.   
In the case of fcc down and hcp up trimers, the reversed (positive) value of $S^{\varphi\varphi}+K^{\varphi\varphi}$ means that 
the ground state of these trimers is rotated by 90\textdegree\ with respect to the conventional, high symmetry directions of the Néel state, see figs.~\ref{fig:geometry}(b) and~\ref{fig:geometry}(c).
Similarly, Gao \etal~\cite{Gao2008} found that triangular Mn islands of different stackings exhibit different easy directions inside the 120\textdegree\ Néel structure.

\section{Mono-layers}

\begin{figure}[!tb]
  \begin{center} \includegraphics[angle=-90,width=\columnwidth]{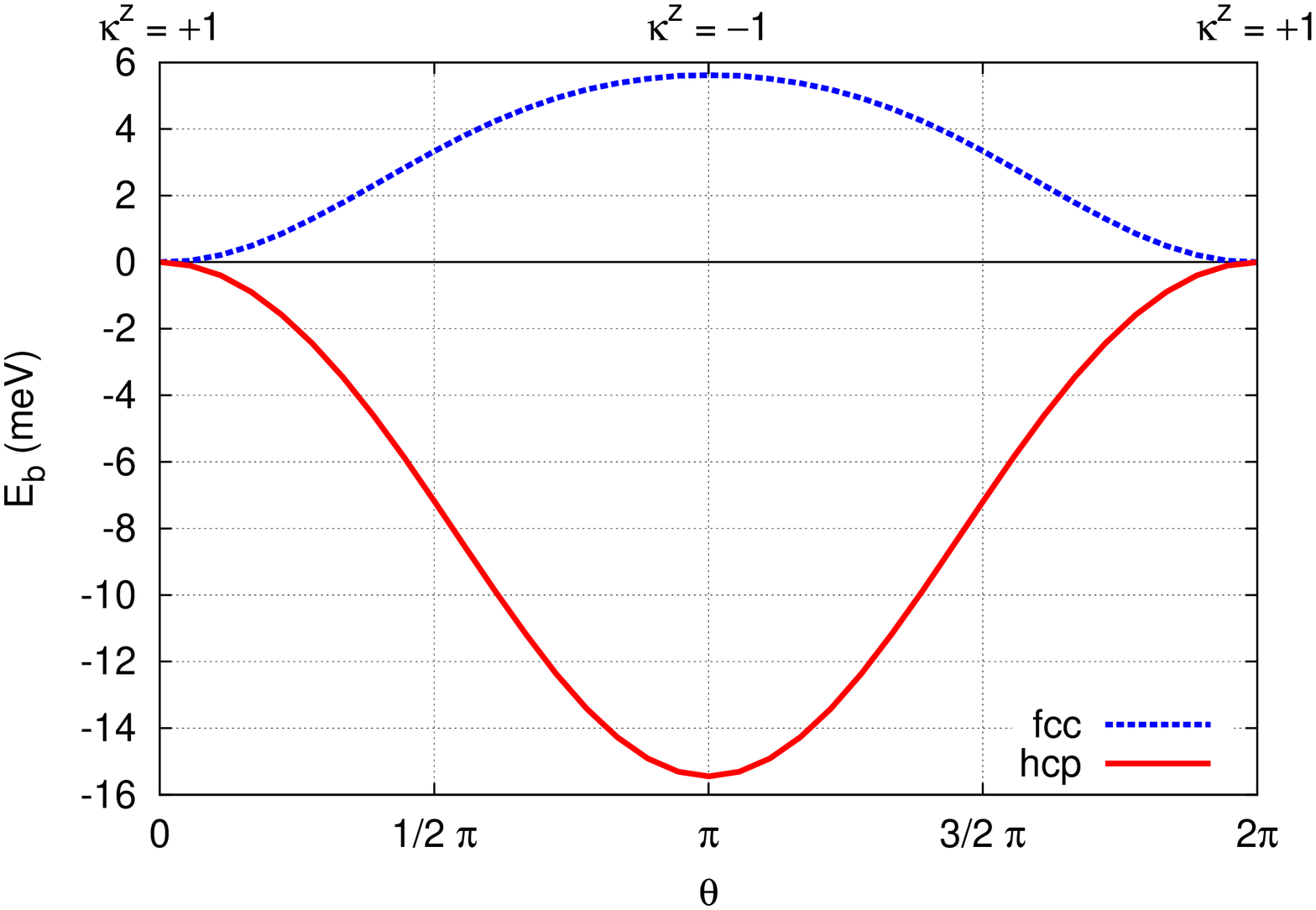} \end{center}
  \caption[]{Rotational energy of the mono-layers about the $y$ axis, $E^{+}_{y}(\vartheta)$.
             The lines were calculated with a resolution of~10\textdegree.
             The $\kappa^z$ component of the chirality vector is indicated above the graph. 
             The energy curves are shifted to zero at the initial configuration, $\vartheta=0$.
             \label{fig:Erotx-ML}
             }
\end{figure}

For both fcc and hcp stacked mono-layers we obtained 
a spin magnetic moment of $3.70\,\mu_\text{B}$ and an orbital magnetic moment of $0.02\,\mu_\text{B}$ for the Cr atoms.
The band energies while rotating the magnetic configuration around the axis lying in the plane of the mono-layer are shown in figure~\ref{fig:Erotx-ML}. 
Using the parameters in table~\ref{tab:parameters} the 
results of the first principles calculations can be fitted with a high accuracy by the function given in 
equation~(\ref{eq:energy-roty}).
Due to the fact that each site in the mono-layer forms a $C_{3}$ symmetry centre, the 
energy of the system contains at best fourth order terms ($\sim \cos(4\varphi)$) in case
of in-plane global rotations. Correspondingly, the band energy turned out to be practically 
independent on the angle of rotation around the $C_{3}$ axis, $S^{\varphi\varphi}+K^{\varphi\varphi}<1\;\mu\text{eV}$.
Note that Szu\-nyogh \etal\ found non-vanishing in-plane anisotropy parameters for the (111) layers in bulk \chem{IrMn_3}~\cite{Szunyogh2009} or at the \chem{IrMn_3/Co} interface~\cite{Szunyogh2011}
since for these systems the above symmetry does not apply.

Fitting the out-of-plane rotational energy in figure~\ref{fig:Erotx-ML} to equation~(\ref{eq:energy-roty}), we obtain nearly the same $S^{zz}-2K^{zz}$ parameters 
for the fcc and hcp mono-layers. These parameters are somewhat reduced in magnitude 
as compared to those for the trimers. 
As obvious from the nearly $\cos\vartheta$-like dependence of the band-energy curves in 
figure~\ref{fig:Erotx-ML}, the out-of-plane rotational energies are dominated by the normal-to-plane components of the DM interactions 
and $D^z$ is opposite in sign for the fcc and the hcp mono-layers.
Note that for the mono-layer case $D^z$ is the
only interaction which distinguishes between the two N\'eel states with opposite chiralities.

The $D^z$ parameters are almost an order larger in magnitude for the mono-layers than for the trimers, see table~\ref{tab:parameters}.
This can be understood due to the following reasoning.
The main contribution to the DM interactions is due to the nearest Cr neighbours. 
Since in the case of mono-layers the number of nearest neighbours is three times larger then those in the trimers, 
a corresponding enhancement of $D^z$ is expected.
More quantitatively, the magnetic unit cell of the mono-layers is composed of 3~up and 3~down elementary trimers. 
Hence, a first estimation of the energy difference between the two chirality states of the mono-layer 
could be 3~times the sum of the energy differences of the trimers.
(It should be recalled that the chirality index of the up and down trimers are opposite in a mono-layer, 
therefore, the chiral energy of the down trimer should be subtracted from that of the up trimer.)
From the data of table~\ref{tab:parameters} we calculate $\Delta{}E_{\text{ML}}/\left(\Delta{}E_{\text{up}}-\Delta{}E_{\text{down}}\right)=5.003$ for the fcc 
mono-layer and 5.118 for the hcp mono-layer. 
The large deviation of these values from~3 indicates that the spin-interactions in a mono-layer are rather different from those in the trimers and/or interactions between more distant pairs have important contributions.

\section{Conclusions}

With regard to the 120\textdegree\ N\'eel structure, we systematically investigated the in-plane, out-of-plane and chiral magnetic anisotropy energies 
of Cr trimers in four different geometries and also of fcc and hcp stacked Cr mono-layers deposited on the Au(111) surface.
Only one out of the four geometric positions for the trimer was considered earlier and here 
we showed that the DM interactions depend intriguingly on the geometry.
The magnetic ground state of the systems turned out to be formed due to an interplay between the DM interactions and the two-site anisotropy.
The actual values of the corresponding parameters determine the energy barrier between the local energy minima related to the magnetic states with
different chiral indices.
Moreover, we revealed an unconventional in-plane easy axis in fcc down and in hcp up trimers.
It should be noted that, in terms of SP-STM experiments, the different magnetic states of otherwise indistinguishable fcc and hcp stacked Mn over-layers 
on Ag(111) became identifiable~\cite{Gao2008}.
The theoretical investigation of the underlying phenomena might thus gain considerable attention.

\ack

Financial support was provided by the Hungarian National Research Foundation (contract No.~OTKA 77771 and 84078)
and in part by the European Union under FP7 Contract No.\ NMP3-SL-2012-281043 FEMTOSPIN.
The work of LS was supported by the European Union, co-financed by the
European Social Fund, in the framework of T\'AMOP 4.2.4.A/2-11-1-2012-0001 National Excellence Program.

\appendix

\section{Magnetic excitation in a 3-sub-lattice antiferromagnet}\setcounter{section}{1}

In order to derive the spin-wave Hamiltonian for the N\'eel antiferromagnetic 
configuration we will follow the method detailed in reference~\citenum{Udvardi2003}. 
The time evolution of the magnetic moments is given by the Landau-Lifshitz equation.~\cite{Landau1935}
In a local coordinate system where the direction of the magnetization in sub-lattice~$p$, $\vect{\sigma}_{p}$, and two transverse unit vectors,
$\vect{e}_p^1$ and $\vect{e}_p^2$, form a right-hand system it has the following form~\cite{Rozsa2014}:
\begin{eqnarray}
    M \frac{\partial\alpha_{pi}}{\partial t} && =  \gamma \frac{\partial E_\text{b}}{\partial \beta_{pi}} \\ 
    M \frac{\partial\beta_{pi}}{\partial t}  && = -\gamma \frac{\partial E_\text{b}}{\partial \alpha_{pi}} ,
\end{eqnarray}
where $E_\text{b}$ is the band energy of the system, $M$ is the magnetization and $\alpha_{pi}$ and $\beta_{pi}$ are the angle of
rotation around the transverse directions $\vect{e}_p^1$ and $\vect{e}_p^2$ at site $i$ of the sub-lattice $p$. Introducing the variables
\begin{equation}
  q_{pi} = \sqrt{\frac{M}{\gamma}}\alpha_{pi} , \qquad
  p_{pi} = \sqrt{\frac{M}{\gamma}}\beta_{pi} ,
\end{equation}
and expanding the energy up to second order around the N\'eel state the Landau-Lifshitz equation has the 
form of~\cite{Udvardi2003}:
\begin{eqnarray}
  \frac{\partial q_{pi}}{\partial t} &&=  \sum_{qj} \left( B_{pi,qj} q_{qj} + A_{pi,qj}p_{qj} \right), \\
  \frac{\partial p_{pi}}{\partial t} &&=  \sum_{qj} \left(-C_{pi,qj} q_{qj} - B_{qj,pi}p_{qj} \right),
\end{eqnarray}
where the $A_{pi,qj}$, $B_{pi,qj}$ and $C_{pi,qj}$ matrices are related to the second derivatives of the
band energy with respect to the transverse change of the exchange field:
\begin{eqnarray}
  \fl A_{pi,qj} = \frac{\gamma}{M} \frac{\partial^2 E_\text{b}}{\partial \alpha_{pi}\partial \alpha_{qj}} , \qquad 
  B_{pi,qj} = \frac{\gamma}{M} \frac{\partial^2 E_\text{b}}{\partial \alpha_{pi}\partial \beta_{qj}} , \nonumber \\
  C_{pi,qj} = \frac{\gamma}{M} \frac{\partial^2 E_\text{b}}{\partial \beta_{pi}\partial \beta_{qj}} .
\end{eqnarray}
The analytic formulas for the second derivatives can be found in reference~\citenum{Balogh2012}.
After applying a lattice Fourier transform in space and a continuous Fourier transform in time 
the magnon frequencies will be the solutions of the following eigenvalue equation:
\begin{equation}
  \begin{pmatrix} u(\vect{k}) \\ v(\vect{k}) \end{pmatrix} = 
  \rmi\omega(\vect{k}) \begin{pmatrix}  \mathbf{B}(\vect{k}) &  \mathbf{A}(\vect{k})    \\
                                       -\mathbf{C}(\vect{k}) & -\mathbf{B}^+(-\vect{k}) \end{pmatrix}
  \begin{pmatrix} u(\vect{k}) \\ v(\vect{k}) \end{pmatrix} , 
\end{equation}
where the $\mathbf{A}(\vect{k})$, $\mathbf{B}(\vect{k})$, and $\mathbf{C}(\vect{k})$ $3\times 3$ 
matrices are the Fourier transform of the corresponding second derivatives, e.g.,
\begin{equation}
 A_{pq}(\vect{k}) = \sum_{j} A_{p0,qj}\rme^{\rmi\vect{kR}_j}
\end{equation}
with $\vect{R}_j$ running over the lattice sites of the $q$-th sub-lattice.

\providecommand{\newblock}{} 
\bibliographystyle{unsrt}          %% Vancouver reference style

\bibliography{cr3}

\end{document}